\documentclass[aps,prl,twocolumn,showpacs]{revtex4}

\usepackage{graphicx}
\usepackage{longtable}
\usepackage{amssymb,amsmath}
\usepackage{setspace}

\begin{document}

\title{Magnetic dipole moment of $^{57,59}$Cu measured by in-gas-cell laser spectroscopy}

\author{T.E.~Cocolios$^1$, A.N.~Andreyev$^1$, B.~Bastin$^1$, N.~Bree$^1$, J.~B\"uscher$^1$, J.~Elseviers$^1$, J.~Gentens$^1$, M.~Huyse$^1$, Yu.~Kudryavtsev$^1$, D.~Pauwels$^1$, T.~Sonoda$^{1,2}$, P.~Van den Bergh$^1$, P.~Van Duppen$^1$}
\affiliation{$^1$ Instituut voor Kern- en Stralingsfysica, K.U. Leuven, B-3001 Leuven, Belgium}
\affiliation{$^2$ RIKEN, Wako, Saitama 305-0198, Japan}

\date{\today}

\begin{abstract}
In-gas-cell laser spectroscopy study of the $^{57,59,63,65}$Cu isotopes has been performed for the first time using the 244.164 nm optical transition from the atomic ground state of copper. The nuclear magnetic dipole moments for $^{57,59,65}$Cu relative to that of $^{63}$Cu have been extracted. The new value for $^{57}$Cu of $\mu(^{57}Cu) = +2.582(7) \mu_N$ is in strong disagreement with the previous literature value but in good agreement with recent theoretical and systematic predictions.
\end{abstract}

\pacs{21.10.Ky, 27.40.+z, 27.50.+e, 42.62.Fi}

\maketitle

 With more than 3000 nuclei known so far, the present nuclear chart offers a vast landscape to study mesoscopic systems. Many of these nuclei cannot be described by \emph{ab initio} calculations and theory uses models based on a fundamental or phenomenological approach in order to describe observables of isotopes yet to discover.
 The confrontation of experimental data with the theoretical predictions does not only allow for fine tuning of theory but also for discovering new aspects of the interactions at work in the atomic nucleus. This is especially the case when studying isotopes with extreme proton-to-neutron ratios.
 In nuclear structure, the identification of the magic numbers $2,8,20,28,50,82,126$ \cite{Goe50} is the foundation for the shell model of the nucleus. 
 While these magic numbers are well established in nuclei close to the valley of $\beta$-stability, their universality is strongly questioned \cite{Sor08}.

Of special interest is the magic number $28$ as it is the smallest magic number issued from the spin-orbit interaction added to the nuclear potential. Both the $N=28$ isotones \cite {Bas07,Gau09} and the nickel ($Z=28$) isotopes \cite{Yur04,Per06} are under intensive investigation to probe their magic character. With $N=Z=28$, $^{56}$Ni is expected to be doubly magic. While it displays a high $2^+_1$ excited state in comparison to the other nickel isotopes \cite{Yur04} and a sudden change in the two-neutron and two-proton separation energies \cite{AME2003}, both characteristic of a doubly magic nucleus, the evolution of the transition strength $B(E2)$ and the behavior of the nuclei in the vicinity point towards particle excitations across the shell gaps and a breaking of this magic core \cite{Kra94,Lis03,Hon04}.

The nuclear magnetic dipole moment is a very sensitive tool to study the nuclear structure in the vicinity of magic nuclei. Indeed, the odd-$A$ $_{29}$Cu isotopes can be described as a single proton coupled to an even-$A$ $_{28}$Ni core and their magnetic dipole moment should in principle be defined by the latter particle only. The copper isotopes have therefore been extensively studied \cite{Rik00,Wei02,Gol04,Min06,Sto08,Sto08b,Fla09}. The magnetic moments from $N=30$ up to $N=40$ depart strongly from the Schmidt moment of a single proton in the $1p_{3/2}$ orbital \cite{Gol04}; this  trend continues while approaching $N=28$. This motivated further studies towards $^{57}$Cu \cite{Min06,Sto08}. 
 Recent shell model calculations using the GXPF1 interaction \cite{Hon04,Sto08b} give a good description of the magnetic moment of the copper isotopes from $N=40$ to $30$ but failed to reproduce the value of $^{57}$Cu \cite{Min06}, the isotope closest to the doubly magic $^{56}$Ni.

Indeed, the $\beta$-NMR measurement reported in \cite{Min06}, made at an in-flight facility, came as a surprise. A magnetic moment $|\mu(^{57}\textrm{Cu})|=2.00(5)\mu_N$ was measured, compared to a predicted value of $2.489\mu_N$ \cite{Hon04,Min06,Sto08b}, pointing towards a more significant shell breaking around $^{56}$Ni compared to what was included in the model.
 Other calculations \cite{Sem96,Gol04} suggested similarly large values for $\mu(^{57}\textrm{Cu})$.
 Note however that the $\beta$-NMR resonance from which the $\mu(^{57}\textrm{Cu})$ is extracted (Fig.~1 in \cite{Min06}) is limited to a single point and has not yet been reproduced. This called therefore for verification using a different radioactive ion beam technique, \emph{e.g.}~laser spectroscopy at an ISOL facility \cite{Sto08}.
 From an experimental point of view, this is a challenging task as the production rate of the $T_{z}=-1/2$ $^{57}$Cu isotope is small and its half-life is short ($T_{1/2}=199$ ms). The in-source laser spectroscopy of radioactive copper isotopes, as developed in high-temperature ISOL target ion source systems \cite{Wei02,Sto08}, is a very sensitive technique but can suffer from significant delay losses. In contrast to this, laser ionisation spectroscopy in a buffer gas cell coupled to an on-line isotope separator allows the study of short-lived isotopes \cite{Son09} providing higher sensitivity and accuracy compared to the high-temperature systems thanks to the smaller total laser line width.
 In this letter, we report about the first successful measurement of the magnetic dipole moment of $^{57}$Cu using in-gas-cell laser spectroscopy. 

The experiment was performed at the Leuven Isotope Separator On-Line (LISOL) facility of the Centre de Recherche du Cyclotron (CRC), Louvain-La-Neuve (Belgium). Beams of $^3$He ($25$ MeV, $2$ $\mu$A) or protons ($30$ MeV, $2$ $\mu$A) impinged on a thin natural nickel target (thickness $5$ $\mu$m) placed in the LISOL dual chamber gas cell \cite{Kud09}. The radioactive isotopes are produced through the reactions $^{58}$Ni(p,2n)$^{57}$Cu, $^{60}$Ni(p,2n)$^{59}$Cu and $^{58}$Ni($^3$He,pn)$^{59}$Cu. The radioactive recoils are stopped and thermalised in $130$ mbar of argon. Stable $^{63,65}$Cu atoms are also produced by the resistive heating of a natural copper filament inside the gas cell.

The atoms are brought towards the ionization chamber of the gas cell by the gas flow where they are ionised to a Cu$^+$ state using a resonant two-step two-color laser ionization process \cite{Kud09,Kud01}. The ions exit the gas cell via a 1 mm exit hole and are caught by a radio-frequency sextupole ion guide before being accelerated to an energy of 40 keV. The beam is further separated according to the isotope mass-to-charge ratio by a dipole magnet. Typical production rates are about 6 ions$\cdot s^{-1}$ for $^{57}$Cu and $1.7\cdot10^4$ or $1.7\cdot10^5$ ions$\cdot s^{-1}$ for $^{59}$Cu using protons or $^3$He, respectively. While scanning the laser frequency, two beams are extracted and counted simultaneously at two different detection stations, $^{57,63}$Cu or $^{59,65}$Cu, respectively. After mass separation, the radioactive isotopes ($^{57,59}$Cu) are implanted in a tape station and counted via their respective $\beta$ decay using three plastic detectors (efficiency $50\%$ \cite{Pau08}) while the stable isotopes ($^{63,65}$Cu,) are simultaneously counted by a channeltron electron multiplier placed after the collector chamber of the mass separator.

\begin{figure}
\includegraphics[width=\columnwidth]{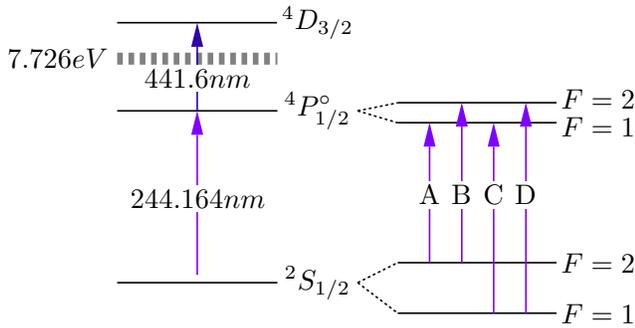}
\caption{\label{fig:scheme}(color online) Laser ionization scheme of copper used in this work. The right part shows the hyperfine splittings and transitions. The thick dashed line is the ionization potential.}
\end{figure}

\begin{figure}
\includegraphics[width=\columnwidth]{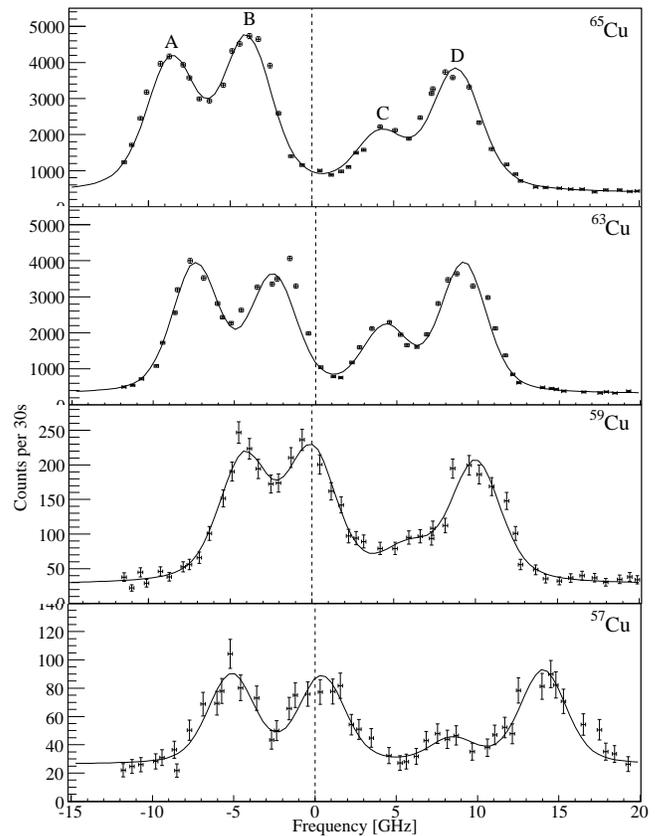}
\caption{\label{fig:spectrum}Typical examples of the single hyperfine spectra of $^{57,59,63,65}$Cu. Each point is sampled for 30 seconds. $^{57}$Cu and $^{63}$Cu are measured simultaneously; so are $^{59}$Cu and $^{65}$Cu. The frequency axis is centered at the center of gravity of $^{63}$Cu.
 The lines are the best fits of four Voigt profiles on top of a constant background, with free amplitudes for each peak, a common full width at half maximum and relative positions constrained by a linear combination of the transition center of gravity and the two hyperfine parameters.}
\end{figure}

The laser spectroscopy is performed by scanning the frequency of the first step laser across the transition from the $3d^{10}4s$ $^2S_{1/2}$ atomic ground state to the $3d^94s4p$ $^4P_{1/2}$ atomic excited state at $244.164$ nm; the ionization scheme is shown in Fig.~\ref{fig:scheme}. The resonances are identified by counting the number of ions extracted as a function of the applied laser frequency. The interaction of the nuclear spin $I=3/2^-$, for all isotopes, and the electronic total angular momentum $J=1/2$, for both atomic levels, yields two sub-levels with quantum numbers $F=1,2$ for each atomic level; the resulting hyperfine structure has four components, as visible in Fig.~\ref{fig:spectrum}. The electronic angular momenta $J_1,J_2=1/2$ restrict the sensitivity of this transition to the magnetic dipole moment. This study can therefore not extract any information on the electric quadrupole moment of the copper isotopes ground states.
The large splitting in both atomic levels allows for the extraction of the hyperfine parameter $A_{hf}$ for each atomic level, as detailed in \cite{Sto08}.

\begin{figure}
\includegraphics[width=\columnwidth]{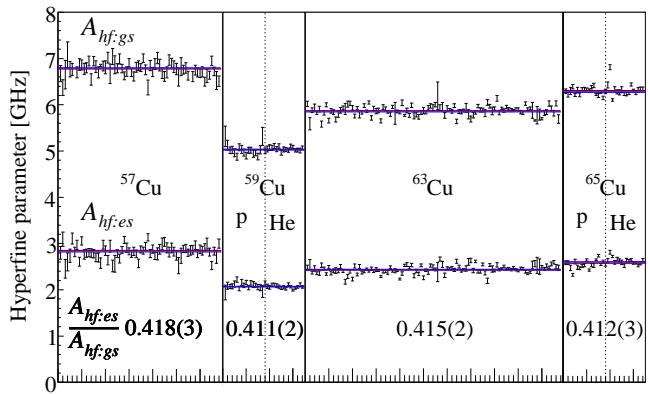}
\caption{\label{fig:sys}(color online) Systematic extracted hyperfine parameters $A_{hf}$ for $^{57,59,63,65}$Cu for the atomic ground state ($A_{hf:gs}$) and the atomic excited state ($A_{hf:es}$). For $^{59,65}$Cu, data using both reactions are presented, identified by the primary beam used, proton (p) or $^3$He (He) respectively. The solid lines are the averages through the points.}
\end{figure}

Each isotopes has been measured repeatedly to ensure the reproducibility of the data. In total, $34$ independent measurements are available for $^{59}$Cu and $^{65}$Cu, $68$ for $^{57}$Cu and $106$ for $^{63}$Cu.
 The hyperfine parameters extracted for every run are consistent to each other and no systematic drift in these parameters has been observed, as shown in Fig.~\ref{fig:sys}. The average value over all the measurements for each of those parameters is shown in Table \ref{tbl:moments}. Off-line, the pressure dependence of the resonance line width and of the center of gravity position were investigated in details and are reported in \cite{Son09}.

The hyperfine parameters for the atomic ground state of $^{63,65}$Cu are known with good accuracy \cite{Fig67} (see table \ref{tbl:moments}). The results from this work, $A_{hf:gs} = 5.858(10)$ GHz and $6.288(17)$ GHz, respectively, are fully consistent with those. Moreover, as shown in Fig.~\ref{fig:sys} and as expected in the absence of hyperfine anomaly, the ratio of the two hyperfine parameters remains constant for all isotopes.
 A nine-fold increase in accuracy is observed for the hyperfine parameter $A_{hf:gs}$ of $^{59}$Cu as given by the present in-gas-cell laser spectroscopy measurement with respect to the high-temperature in-source laser spectroscopy work \cite{Sto08}. This is due to the improved total resonance line width ($3.5$ GHz vs.~$4.5$ GHz), the larger separation of the hyperfine levels of the $3d^94s4p$ $^4P_{1/2}$ level compared to the $3d^{10}4p$ $^2P_{1/2}$ level and the high number of independent measurements.
 Supported by the good agreement on the stable isotopes, the consistency of the ratio of the two hyperfine parameters and based on the precise knowledge of the magnetic moment of $^{63}$Cu \cite{Lut78,Sto05}, the moments of $^{57,59,65}$Cu are extracted, as detailed in \cite{Sto08}, from both atomic levels. The results are given in Table \ref{tbl:moments}. The signs are determined based on the ordering of the peaks considering the relative intensity of the $F=1\rightarrow1$ transition (labeled C), much lower with respect to the others, as seen in Fig.~\ref{fig:spectrum}.

\begin{table*}
\caption{\label{tbl:moments}Measured hyperfine parameters $A_{hf:exp}$ for the atomic ground (gs) and excited (es) states and the deduced moments $\mu_{exp}$ using $^{63}$Cu as the reference isotope. The literature values $A_{hf:lit:gs}$ \cite{Sto08,Fig67}, $\mu_{lit}$ \cite{Lut78,Gol04,Sto05,Min06} and theoretical calculations using GXPF1 \cite{Hon04,Sto08b} are given for comparison; no literature is available on the atomic excited hyperfine parameter.}
\begin{ruledtabular}
\begin{tabular}{cccccccc}
$A$ & $I$ & $A_{hf:exp:gs}$ [GHz] & $A_{hf:lit:gs}$ [GHz] & $A_{hf:exp:es}$ [GHz] & $\mu_{exp}$ $[\mu_N]$ & $\mu_{lit}$ $[\mu_N]$ & $\mu_{GXPF1}$ $[\mu_N]$\\
\hline
$57$ & $3/2^-$ & $6.785(15)$ & - & $2.834(16)$ & $+2.582(7)$ & $2.00(5)$ & $2.489$\\
$59$ & $3/2^-$ & $5.033(10)$ & $4.87(9)$ & $2.069(8)$ & $+1.910(4)$ & $+1.891(9)$ & $1.886$\\
$63$ & $3/2^-$ & $5.858(10)$ & $5.866908706(20)$ & $2.432(8)$ & - & $2.2273602(13)$ & $2.251$\\
$65$ & $3/2^-$ & $6.288(17)$ & $6.284389972(60)$ & $2.588(15)$ & $+2.387(7)$ & $2.3818(3)$ & $2.398$\\
\end{tabular}
\end{ruledtabular}
\end{table*}

Good agreement is found with previous moment measurements of the $^{59,65}$Cu isotopes. The measured moment of the lightest isotope $^{57}$Cu ($\mu=+2.582(7)\mu_N$) displays however a major difference with the literature value ($|\mu|=2.00(5)\mu_N$) \cite{Min06}. A careful inspection of our running conditions and of our analysis has been performed. Moreover, the systematic measurement of $^{63}$Cu, the high reproducibility of the spectra and the good agreement of each measured isotope with the established literature values confirm the accuracy of the method. The literature value in \cite{Min06} is therefore questioned.

\begin{figure}
\includegraphics[width=\columnwidth]{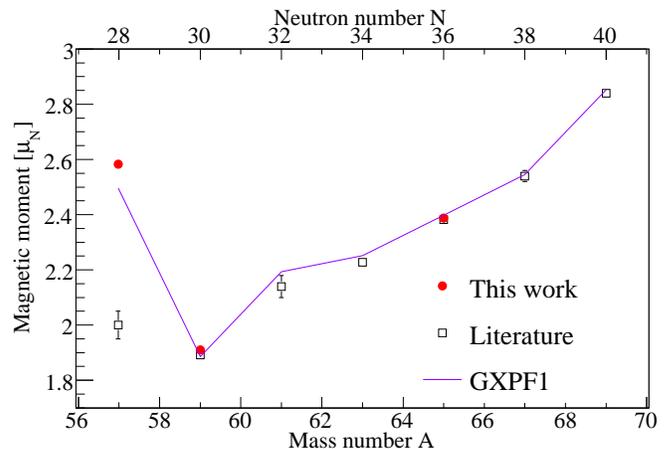}
\caption{\label{fig:moments}(color online) Ground state nuclear magnetic dipole moments of the odd-$A$ copper isotopes. The full circles are the moments published in this Letter, the full squares are the experimental values from the literature \cite{Rik00,Gol04,Min06,Lut78,Sto05}. 
 Theoretical calculations using the GXPF1 interaction, a $^{40}$Ca core and the full $fp$-shell valence space \cite{Hon04,Sto08b} is shown with a solid line. The Schmidt value $\mu_{Schmidt}=+3.79\mu_N$ falls out of the range of the figure.}
\end{figure}

All the magnetic moments of the copper isotopes depart strongly from the Schmidt value $\mu_{Schmidt}=+3.79\mu_N$ of a single proton in a $1p_{3/2}$ orbital. In the case of the semi-magic $^{69}$Cu$_{40}$ nucleus, the difference between the Schmidt moment and the experimental moment is very well reproduced by the shell-model calculation from Golovko \emph{et al.}~\cite{Gol04} ($\mu=+2.87(13)\mu_N$). $^{68}$Ni was taken as a closed-shell core but including effects of core polarisation, meson exchange current, $\Delta$-isobars and relativistic corrections in perturbation theory. The same calculations for $^{57}$Cu$_{28}$ using $^{56}$Ni as the core give $\mu=+2.40(18)\mu_N$ and reproduce the new measured value.

Theoretical studies considering a $^{40}$Ca core and the full $fp$-shell valence space are also in agreement with the dipole moment of $^{57}$Cu as measured in this Letter, predicting a magnetic moment of $\mu=+2.48\mu_N$ using the FPD6 interaction \cite{Sem96} or $\mu=+2.489\mu_N$ using the GXPF1 interaction (with effective $g$ factors $g^{eff}_s = 0.9g^{free}_s$, $g^{free}_l = 1.1$ for protons and $g^{free}_l = -0.1$ for neutrons) \cite{Hon04,Sto08b}. The moments of the isotopes between $N=28$ and $N=40$ have also been extracted with the latter interaction and reproduce the experimental data accurately (see Fig.~\ref{fig:moments}).

The new value for the magnetic dipole moment of $^{57}$Cu can also be used together with the one of its mirror partner $^{57}$Ni ($-0.7975(14)\mu_N$ \cite{Oht96}) to extract the isoscalar spin expectation value $\big\langle \sum \sigma_Z \big\rangle = 0.75(2)$ according to the formalism described in \cite{Min06}.  This quantity reflects the contribution from the nucleon spin to the magnetic moment. Our value is in strong disagreement with the value of $-0.78(13)$ from \cite{Min06}. However, it is in reasonable agreement with the calculated values $0.71$ using the FPD6 interaction \cite{Sem96} and $0.51$ using the GXPF1 interaction \cite{Hon04,Min06}. The departure of this value from 1 is an extra indication of a non-pure $p_{3/2}$ nuclear configuration (see Fig.~3 in \cite{Min06}).

Moreover, the dipole moments of $^{57}$Cu can be estimated based on its respective mirror nucleus $^{57}$Ni \cite{Buc01,Buc83}. The deduced moment $\mu=+2.49(3)\mu_N$ is again in agreement with our measurement. 
 The magnetic moment of $^{57}$Cu and $^{57}$Ni can also be combined to calculate the magnetic dipole moment of $^{58}$Cu according to the additivity rule \cite{Sto08}. A value of $\mu=+0.595(2)\mu_N$ is found, in agreement with the experimental value $+0.52(8)\mu_N$ \cite{Sto08}.

Our work brings out how good the GXPF1 interaction describes the structure near $^{56}$Ni as proven by the very sensitive reproduction of the magnetic dipole moment of the chain $^{57-69}$Cu. There is indeed no need for a more significant shell breaking than introduced in \cite{Hon04}, unlike stated previously in \cite{Min06}.

To conclude, we have reported the first on-line magnetic moment measurement of an exotic isotope using in-gas-cell resonant ionization laser spectroscopy coupled to a mass separator. The system is proven to be very stable, has a superior accuracy compared to high-temperature in-source laser spectroscopy due to a lower total resonance line width. Furthermore, it allows laser spectroscopy measurements of short-lived radioactive isotopes and of isotopes from refractory elements that are not possible using high-temperature target-ion source systems.
 This new technique opens therefore exciting possibilities for the future radioactive ion beam facilities across the world making use of the gas-cell technology (e.g.~GANIL, NSCL, RIKEN).

 The hyperfine parameter of the $3d^94s4p$ $^4P_{1/2}$ level in copper has been measured for the first time. Moreover, the known magnetic moments for $^{59}$Cu and $^{65}$Cu are well reproduced. The discrepancy with the $\beta$-NMR measurement of $^{57}$Cu questions however the correctness of the value published in \cite{Min06}. Finally, a good agreement of the new measurement with recent theoretical calculations and with the prediction from the mirror nucleus $^{57}$Ni is found. 

Besides, other isotopes displaying large hyperfine splittings are very well suited for this type of measurement. The neutron-deficient silver, indium and tin isotopes, approaching $N=50$, are expected to possess large magnetic dipole moments. They are therefore ideal to probe this shell closure. The in-gas-cell laser spectroscopy technique can also be improved by reducing the resonance line width further, performing the laser spectroscopy in a Laser Ion Source Trap (LIST) as recently shown in \cite{Son09}.

\begin{acknowledgments}
We thank the CRC team, Louvain-La-Neuve (Belgium). This work was supported by FWO-Vlaanderen (Belgium), GOA/2004/03 (BOF-K.U.Leuven), the IUAP - Belgian State  Belgian Science Policy - (BriX network P6/23) and by the European Commission within the Sixth Framework Programme through I3-EURONS (Contract RII3-CT-2004-506065).
\end{acknowledgments}


\end{document}